\documentclass{PoS}
\usepackage{graphicx}
\usepackage{amsmath,bm}

\DeclareMathOperator{\tr}{tr}

\title{The effects of explicit chiral symmetry breaking multiquark interactions on the spin 0 and 1 meson nonets: the ruling of the vector mesons.}

\ShortTitle{The effects of E$\chi$SM multiquark interactions on the spin 0 and 1 meson nonets: the ruling of the vector mesons.}

\author{\speaker{Jorge Morais}\thanks{Based on the talk given at Hadron 2017, Salamanca. Work supported in part by Funda\c{c}\~{a}o para a Ci\^{e}ncia e a Tecnologia (FCT), through PhD grant SFRH/BD/110315/2015 and project UID/FIS/04564/2016, and Centro de F\'{i}sica da Universidade de Coimbra.}\\
        CFisUC, Department of Physiscs, University of Coimbra, 3004-516 Coimbra, Portugal\\
        E-mail: \email{jorge.m.r.morais@gmail.com}}

\author{Brigitte Hiller\\
        CFisUC, Department of Physiscs, University of Coimbra, 3004-516 Coimbra, Portugal\\
        E-mail: \email{brigitte@teor.fis.uc.pt}}
    
\author{Alexander A. Osipov\\
    	JINR, Bogoliubov Laboratory of Theoretical Physics, 141980 Dubna, Russia\\
    	E-mail: \email{osipov@nu.jinr.ru}}

\abstract{We have recently extended the scalar-pseudoscalar sector of a generalized NJL Lagrangian that includes all NLO non derivative interactions in Nc counting (including explicit symmetry breaking ones) in order to incorporate the spin 1 mesons in the low-lying ground state of QCD \cite{Morais17}. Upon bosonization, the well known mixing of the scalar-vector and of the pseudoscalar- axial-vector fields occurs in the quadratic part of the Lagrangian. We show that a linearized diagonalization of these terms can be effected in a completely general way without compromising the underlying symmetries of the Lagrangian \cite{Morais17b}. The resulting spin 1 mass spectra evidence a relation involving only the vector and axial-vector meson masses and the constituent quark masses. We discuss the dominant role of this relation in the fits and we show that the model may be fitted to accommodate to a very good accuracy the 4 low-lying meson spectra.}

\FullConference{XVII International Conference on Hadron Spectroscopy and Structure - Hadron2017\\
		25-29 September, 2017\\
		University of Salamanca, Salamanca, Spain}

\begin{document}

\section{Introduction - Spin 0 Model}

Under a reasonable assumption that chiral symmetry and its breaking mechanisms constitute predominant features of low-energy QCD, an idea which dates back to the Nambu--Jona-Lasinio (NJL) model \cite{NJL}, we discuss a 3 flavor NJL-type model which includes all non-derivative effective multiquark vertices (involving products of spin 0 bilinears) which are relevant for the dynamical breaking of chiral symmetry in 4 dimensions \cite{Osipov13,Andrianov}. We argue that this set coincides with a next-to-leading-order (NLO) expansion in $N_c^{-1}$, with the usual 4 quark NJL vertex and the Dirac mass term constituting the leading order terms. The 6 quark 't Hooft determinant, which is $N_c^{-1}$ suppressed, must be included for 3 flavors in order to explicitly break the $U\left(1\right)_A$ symmetry. Two 8 quark effective vertices, which are of the same order in $N_c$ counting as the 't Hooft term and are required in order for the effective potential to be bounded from below, complete the set of chiral terms.

Owing to the considerable difference between the strange quark mass and that of the lighter quarks, explicit symmetry breaking (ESB) effects are expected to significantly contribute to the dynamics. Consistency with the expansion order demands that we go beyond a simple Dirac mass term and include all ESB terms which might contribute to NLO as well. We do this by letting the quark fields interact with a scalar source and then systematically write all such possible terms up to the relevant order, later identifying this source with a quark current mass matrix.

The model is non-renormalizable and depends explicitly on an energy scale $\Lambda$ which is assumed to be of the order of the chiral symmetry breaking scale $\Lambda_{\chi SB} \sim 1 \, \text{GeV}$. Defining $s_a = \bar{q} \lambda_a q$, $p_a = \bar{q} i \gamma_5 \lambda_a q$, the $U\left(3\right)$ valued field $\Sigma = \frac{1}{2} \left(s_a - i p_a\right)\lambda_a$, an external source $\chi$ which is assumed to transform as $\Sigma$, and scaling each term by the appropriate power of $\Lambda$, we have the chirally symmetric terms
\begin{align}
	\label{main0}
	\mathcal{L}_{int} & = \frac{\bar{G}}{\Lambda^2} \tr{\left(\Sigma^\dagger \Sigma\right)} + \frac{\bar{\kappa}}{\Lambda^5} \left(\det{\Sigma} + \det{\Sigma^\dagger}\right) \nonumber \\
	& + \frac{\bar{g}_1}{\Lambda^8} \left(\tr{\Sigma^\dagger \Sigma}\right)^2 + \frac{\bar{g}_2}{\Lambda^8} \tr{\left(\Sigma^\dagger \Sigma \Sigma^\dagger \Sigma\right)} ,
\end{align}

\noindent and the ESB terms
\begin{align}
	\label{0esb}
	\mathcal{L}_0 & = -\tr{\left(\Sigma^\dagger \chi + \chi^\dagger \Sigma\right)}, & \mathcal{L}_5 & = \frac{\bar{g}_5}{\Lambda^4} \tr{\left(\Sigma^\dagger \chi \Sigma^\dagger \chi\right)} + h.c., \nonumber \\
	\mathcal{L}_2 & = \frac{\bar{\kappa}_2}{\Lambda^3} \epsilon_{ijk} \epsilon_{mnl} \Sigma_{im} \Sigma_{jn} \chi_{kl} + h.c., & \mathcal{L}_6 & = \frac{\bar{g}_6}{\Lambda^4} \tr{\left(\Sigma^\dagger \Sigma \chi^\dagger \chi\right)} + h.c., \nonumber \\
	\mathcal{L}_3 & = \frac{\bar{g}_3}{\Lambda^6} \tr{\left(\Sigma^\dagger \Sigma \Sigma^\dagger \chi\right)} + h.c., & \mathcal{L}_7 & = \frac{\bar{g}_7}{\Lambda^4} \left(\tr{\Sigma^\dagger \chi} + h.c.\right)^2, \nonumber \\
	\mathcal{L}_4 & = \frac{\bar{g}_4}{\Lambda^6} \tr{\left(\Sigma^\dagger \Sigma\right)} \tr{\left(\Sigma^\dagger \chi\right)} + h.c., & \mathcal{L}_8 & = \frac{\bar{g}_8}{\Lambda^4} \left(\tr{\Sigma^\dagger \chi} - h.c.\right)^2 .
\end{align} 

This spin 0 version of the model has been very successful in reproducing the low-lying scalar and pseudoscalar meson spectra along with several strong and radiative meson decays. It has also been applied to studying the QCD phase diagram and equation of state, as well as assessing the prospect of strange quark matter \cite{SQM}.

\section{Inclusion of Spin 1 Mesons}
Owing to its theoretical appeal and phenomenological success, an extension to include spin 1 mesons is a natural step forward. A major motivation for this is the possibility to study decay processes and interactions involving vector and axial-vector mesons. Also, it may provide a means to assess the VMD hypothesis. Additionally, it has been shown in different approaches that the vector mesons have an important impact in the equation of state of QCD \cite{Benic}, a fact which could be studied within our framework as well.

We define the quark bilinears $v_a^{\mu} = \bar{q} \gamma^{\mu}\lambda_a q$,  $a_a^{\mu} = \bar{q} \gamma^{\mu} \gamma_5 \lambda_a q$, and the $U\left(3\right)$ valued fields $R^{\mu},L^{\mu} = \frac{1}{2} \left(v_a^{\mu} \pm a_a^{\mu}\right)\lambda_a$. The symmetry criteria lead to the inclusion of the new terms \cite{model2}:
\begin{align}
	\mathcal{L}'_1 & = \frac{\bar{w}_1}{\Lambda^2} \tr{\left(R^{\mu}R_{\mu} + L^{\mu}L_{\mu}\right)}, & \mathcal{L}'_2 & = \frac{\bar{w}_2}{\Lambda^8} \left[\tr{\left(R^{\mu}R_{\mu} + L^{\mu}L_{\mu}\right)}\right]^2, \nonumber \\
	\mathcal{L}'_3 & = \frac{\bar{w}_3}{\Lambda^8} \left[\tr{\left(R^{\mu}R_{\mu} - L^{\mu}L_{\mu}\right)}\right]^2, & \mathcal{L}'_4 & = \frac{\bar{w}_4}{\Lambda^8} \tr{\left(R^{\mu}R^{\nu}R_{\mu}R_{\nu} + L^{\mu}L^{\nu}L_{\mu}L_{\nu}\right)}, \nonumber \\
	\mathcal{L}'_5 & = \frac{\bar{w}_5}{\Lambda^8} \tr{\left(R^{\mu}R_{\mu}R^{\nu}R_{\nu} + L^{\mu}L_{\mu}L^{\nu}L_{\nu}\right)}, & \mathcal{L}'_6 & = \frac{\bar{w}_6}{\Lambda^8} \tr{\left(R^{\mu}R_{\mu} + L^{\mu}L_{\mu}\right)} \tr{\left(\Sigma^\dagger \Sigma\right)}, \nonumber \\
	\mathcal{L}'_7 & = \frac{\bar{w}_7}{\Lambda^8} \tr{\left(\Sigma^\dagger L^{\mu} \Sigma R_{\mu}\right)}, & \mathcal{L}'_8 & = \frac{\bar{w}_8}{\Lambda^8} \tr{\left(\Sigma^\dagger \Sigma R^{\mu} R_{\mu} + \Sigma \Sigma^\dagger L^{\mu} L_{\mu}\right)}, \nonumber \\
	\mathcal{L}'_9 & = \frac{\bar{w}_9}{\Lambda^6} \tr{\left(R^{\mu}R_{\mu} + L^{\mu}L_{\mu}\right)} \tr{\left(\Sigma^\dagger \chi + \Sigma \chi^\dagger\right)}, & \mathcal{L}'_{10} & = \frac{\bar{w}_{10}}{\Lambda^6} \tr{\left(\chi^\dagger L^{\mu} \Sigma R_{\mu} + \Sigma^\dagger L^{\mu} \chi R_{\mu}\right)}, \nonumber \\
	\mathcal{L}'_{12} & = \frac{\bar{w}_{12}}{\Lambda^4} \tr{\left(\chi^\dagger L^{\mu} \chi R_{\mu}\right)}, & \mathcal{L}'_{13} & = \frac{\bar{w}_{13}}{\Lambda^4} \tr{\left(\chi^\dagger \chi R^{\mu} R_{\mu} + \chi \chi^\dagger L^{\mu} L_{\mu}\right)}, \nonumber
\end{align}
\begin{equation}
\mathcal{L}'_{11} = \frac{\bar{w}_{11}}{\Lambda^6} \tr{\left[\left(\Sigma^\dagger \chi + \chi^\dagger \Sigma\right) R^{\mu} R_{\mu} + \left(\Sigma \chi^\dagger + \chi \Sigma^\dagger\right) L^{\mu} L_{\mu} \right]}.
\end{equation}
\noindent Although 13 new parameters are introduced, only some will contribute to the vacuum properties of the model.

The quark Lagrangian is bosonized in a functional integral formalism, introducing the bosonic degrees of freedom $\sigma$ (scalar), $\phi$ (pseudoscalar), $V^{\mu}$ (vector) and $A^{\mu}$ (axial vector). The system is described in the Nambu-Goldstone phase by defining the scalar fluctuations by $\sigma \to \sigma + M$, with $M$ being interpreted as a constituent quark mass matrix. Auxiliary degrees of freedom $s$, $p$, $v$, $a$ are integrated out using a stationary phase approximation (SPA), while the remaining quark determinant is performed using a generalized heat kernel technique.

Upon bosonization, bilinear mixing terms appear of the form 
\begin{equation}
\label{mixing}
\tr{\left(i \left[V^{\mu},M\right]\partial_{\mu}\sigma - \left\lbrace A^{\mu},M \right\rbrace \partial_{\mu}\phi \right)} .
\end{equation}

\noindent Due to the inclusion of higher-order effective vertices, mass diagonalization requires general redefinitions of the fields. The simplest choice consists of the linear shifts
\begin{align}
	\label{shifts}
	& V_{\mu}\rightarrow V'_{\mu} + X_{\mu}, & X_\mu = k\circ\Delta_M\circ\partial_\mu\sigma, & & \left(\Delta_M\right)_{ij} = -i \left(M_i - M_j\right), \\
	& A_{\mu}\rightarrow A'_{\mu} + Y_{\mu}, & Y_\mu=k'\circ\Sigma_M\circ \partial_\mu\phi, & & \left(\Sigma_M\right)_{ij} = M_i + M_j,
\end{align}
\noindent where $k$, $k'$ are adjustable coefficient matrices, and $\circ$ denotes the Hadamard product. Necessarily, the mass eigenfields $V'_{\mu}$ and $A'_{\mu}$ obey new chiral transformation laws which, nonetheless, have been shown to preserve the chiral group structure \cite{Morais17b}. Through this mixing scheme, additional contributions to spin 0 kinetic terms  arise and, consequently, new normalization factors for these fields appear which depend on spin 1 parameters. In turn, the squared masses of the bosons are extracted from the diagonalized Lagrangian in the isospin limit, revealing a full set of linear expressions of the form $ M_{s,p}^2 = c_1 M_{v,a}^2 + c_2$ which relate homologous spin 0 and spin 1 masses in a systematic fashion, with slightly modified versions for the mixed neutral channels ($f_0,\sigma$ and $\eta, \eta'$) which involve combinations of the corresponding spin 1 masses parametrized by mixing angles $\psi_{\sigma,\phi}$. The coefficient $c_1$ consists of a specific combination of the model's parameters of each case, while $c_2$ is either $4 M_i^2$ ($i = u,d,s$) in the scalar case or $0$ in the pseudoscalar case. These relations manifest the underlying strict symmetry constraints which are at the core of the model's conception and give strength to the fact that, although the parameters may be numerous, by no means do they leave space for much arbitrariness in the observables it sets out to describe.

\section{Parameter Fitting}

Parameters $w_2$ to $w_5$ do not contribute to the vacuum properties of the model, effectively lowering the number of new parameters from 13 to 9. Furthermore, parameters $w_1$, $w_6$ and $w_9$ appear everywhere in a fixed combination which can be treated as a single parameter by setting $w_6 = w_9 = 0$. Among the new parameters, $w_7$, $w_{10}$ and $w_{12}$ have been shown to be tightly constrained by mass differences between spin 1 chiral partner mesons. The following 3 relations can be derived within the model:
\begin{align}
2\left(M_{K^*}^2 - \frac{3}{2} \left(M_u - M_s\right)^2\right)^{-1} + 2\left(M_{K_1}^2 - \frac{3}{2} \left(M_u + M_s\right)^2\right)^{-1} = \nonumber \\
M_{\rho}^{-2} + M_{\varphi}^{-2} + \left(M_{a_1}^2 - 6 M_u^2\right)^{-1} + \left(M_{f_1}^2 - 6 M_s^2\right)^{-1} ,\nonumber\\
M_{a_1}^2=\frac{6 M_{u}^4}{M_{u}^2-\varrho^2f_{\pi}^2}, \qquad
M_{K_1}^2=\frac{\frac{3}{2}\left(M_{u}+M_s\right)^4}{\left(M_{u}+M_s\right)^2-4\varrho^2f_K^2},
\end{align}

\noindent This set of relations is particularly significant, because it provides a way do directly determine $M_u$, $M_s$ and $\Lambda$ (which appears in $\varrho$) from observables (spin 1 masses and weak decay constants $f_{\pi}$ and $f_K$).

\begin{table}
	\caption{Empirical input (values in MeV).}
	\label{emp_input}
	\centering
	\begin{tabular}{cccccccccc}
		\hline
		\textbf{$M_\pi$} 
		&\textbf{$M_K$} 
		&\textbf{$M_\eta$}
		&\textbf{$M_{\eta'}$}
		&\textbf{$M_\sigma$}
		&\textbf{$M_\kappa$}
		&\textbf{$M_{a_0}$}
		&\textbf{$M_{f_0}$}
		&\textbf{$M_\rho$}
		&\textbf{$M_{K^*}$}
		\\ 
		\hline
		138
		& 496 
		& 548 
		& 958 
		& 500 
		& 850 
		& 980 
		& 980 
		& 778 
		& 893 \\
		\hline
		\hline
		\textbf{$M_\varphi$}
		&\textbf{$M_{a_1}$}
		&\textbf{$M_{K_1}$}
		&\textbf{$M_{f_1}$}
		&\textbf{$m_u$}
		&\textbf{$m_s$}
		&\textbf{$f_\pi$}
		&\textbf{$f_K$}
		&\textbf{$\theta_\phi$} & \\
		\hline
		1019
		& 1270 
		& 1274 
		& 1426 
		& 4 
		& 100 
		& 92 
		& 111 
		& -15$^{\circ}$ &
		\\
		\hline
	\end{tabular}
\end{table}

\begin{table}
	\caption{Fit results (values of $M_i$, $\Lambda$ in MeV).}
	\label{fit_results}
	\centering
	\begin{tabular}{cccccccc}
		\hline
		\textbf{$\theta_\sigma$}
		& \textbf{$\Lambda$} 
		& \textbf{$M_u$}
		& \textbf{$M_s$}
		& \textbf{$\bm{w_1}$} 
		& \textbf{$\bm{w_6}$} 
		& \textbf{$\bm{w_9}$}
		& \textbf{$\bm{w_{13}}$} 
		\\
		\hline
		25.1$^{\circ}$
		& 1633
		& 244
		& 508
		& -10 
		& 0
		& 0
		& 0\\
		\hline
	\end{tabular}
\end{table}

\begin{table}
	\caption{Values of non-zero parameters in natural units (with $S$ being the scale factor) \cite{Morais17}.}
	\label{natural}
	\centering
	\begin{tabular}{ccccccccc}
		\hline
		\text{c}
		& \textbf{$G$} 
		& \textbf{$\kappa$} 
		& \textbf{$g_1$} 
		& \textbf{$g_2$} 
		& \textbf{$\kappa_2$} 
		& \textbf{$g_3$} 
		& \textbf{$g_4$} 
		& \textbf{$g_5$} \\ 
		\hline
		\text{S}
		& $\frac{f^2\Lambda^2}{M^2}$
		& $\frac{f^4\Lambda^4}{M^3}$
		& $\frac{f^6\Lambda^6}{M^4}$
		& $\frac{f^6\Lambda^6}{M^4}$
		& $\frac{f^2\Lambda^2}{M}$
		& $\frac{f^4\Lambda^4}{M^2}$
		& $\frac{f^4\Lambda^4}{M^2}$
		& $f^2\Lambda^2$ \\
		\text{$\bar c$}
		&  1.0
		& -0.1 
		&  0.05
		& -0.1 
		&  0.01 
		& -1.3 
		&  0.3 
		& -0.5 \\ 
		\hline
		\hline
		\textbf{$g_6$}
		& \textbf{$g_7$} 
		& \textbf{$g_8$} 
		& \textbf{$w_1$} 
		& \textbf{$w_7$} 
		& \textbf{$w_8$} 
		& \textbf{$w_{10}$} 
		& \textbf{$w_{11}$} 
		& \textbf{$w_{12}$} \\
		\hline
		$f^2\Lambda^2$
		& $f^2\Lambda^2$
		& $f^2\Lambda^2$ 
		& $f^2$ 
		& $\frac{f^6\Lambda^4}{M^2}$
		& $\frac{f^6\Lambda^4}{M^2}$
		& $f^4\Lambda^2$
		& $f^4\Lambda^2$
		& $f^2M^2$ \\
		-2.6
		& -0.7 
		& -0.5 
		& -0.1 
		& -0.1 
		&  0.1 
		& -0.5 
		&  0.3 
		& -0.8 \\
		\hline
		
	\end{tabular}
\end{table}

The empirical data in table \ref{emp_input} has been used to fit the model's parameters, and the results are shown in tables \ref{fit_results} and \ref{natural} (the effective couplings in table \ref{fit_results} have been externally fixed without loss of generality). Results in table \ref{natural} are shown in natural units ($\bar{c} = S c$) after a methodical removal of the relevant scales \cite{Manohar,Morais17}, including $\Lambda$ (which estimates the scale of spontaneous symmetry breaking), $M$ (which is characteristic of chirality violations at the quark vertices), and $f$ (which governs the dynamics of the pseudo-Goldstone bosons). It should be noted that the model has a high degree of degeneracy among parameter sets which yield the same vacuum results, but this is expected to be lifted upon introducing thermodynamic parameters.

The resulting $\theta_{\sigma}$, $\Lambda$, and $M_i$ are all obtained within reasonable values. The inclusion of spin 1 mesons leads to a $\Lambda$ which is roughly double of what it had been previously obtained with the spin 0 version, but still within $\mathcal{O}\left(1 \text{GeV}\right)$. The difference $M_s - M_u$ is also enhanced when compared to the previous case. 

The model has been successfully fitted to reproduce the full low-lying meson nonets' spectra. It comes out very explicitly that the model's parameters are very tightly constrained by the symmetries underlying its construction. These constraints are very well apparent in the mass relations between spin 0 and spin 1 mesons. Future prospects include the computation of meson decays, the equation of state and the study of the phase diagram.


\begin{thebibliography}{99}
\bibitem{Morais17} J. Morais, B. Hiller, A. A. Osipov,
\emph{Masses of the lowest spin-0 and spin-1 meson nonets: explicit symmetry breaking effects},
\emph{Phys. Rev. D} {\bf 95}, 074033 (2017)
[{\tt hep-th/1702.06894}]
	
\bibitem{Morais17b} J. Morais, B. Hiller, A. A. Osipov,
\emph{A general framework to diagonalize vector--scalar and axial-vector--pseudoscalar transitions in the effective meson Lagrangian},
\emph{Phys. Lett. B} {\bf 773}, pp. 277-282 (2017)
[{\tt hep-th/1705.04644}]

\bibitem{NJL} Y. Nambu, G. Jona-Lasinio
\emph{Dynamical Model of Elementary Particles Based on an Analogy with Superconductivity I}
\emph{Phys. Rev.} {\bf 122}, 345 (1961)
Y. Nambu, G. Jona-Lasinio
\emph{Dynamical Model of Elementary Particles Based on an Analogy with Superconductivity II}
\emph{Phys. Rev.} {\bf 124}, 246 (1961)

\bibitem{Osipov13} A. A. Osipov, B. Hiller, A. H. Blin,
\emph{Light quark masses in multi-quark interactions},
\emph{Eur. Phys. J. A} {\bf 49:}14 (2013)
[{\tt hep-th/1206.1920}] \\
A. A. Osipov, B. Hiller, A. H. Blin,
\emph{Effective multi-quark interactions with explicit breaking of chiral symmetry},
\emph{Phys. Rev. D} {\bf 88} 054032 (2013)
[{\tt hep-th/1309.2497}]

\bibitem{Andrianov} A. A. Andrianov, V. A. Andrianov,
\emph{Structure of effective fermion models in symmetry-breaking phase},
\emph{Int. J. Mod. Phys. A} {\bf 08}, 1981 (1993)

\bibitem{SQM} J. Moreira, J. Morais, B. Hiller, A. A. Osipov, A. H. Blin,
\emph{Strange quark matter in the presence of explicit symmetry breaking interactions}, \emph{Phys. Rev. D} {\bf 91}, 116003 (2015)
[{\tt hep-th/1409.0336}]

\bibitem{Benic} S. Benic
\emph{Heavy hybrid stars from multi-quark interactions}
\emph{The European Physical Journal A} {\bf 50}, 111 (2014)

\bibitem{Manohar} A. Manohar, H. Georgi,
\emph{Chiral quarks and the non-relativistic quark model},
\emph{Nucl. Phys. B} {\bf 234}, 189 (1984)

\end{thebibliography}
\end{document}